\documentstyle[epsfig,prc,preprint,aps]{revtex}

\tighten

\begin{document}
\draft
 
\begin{title}
{$\bf \Xi^{-} d\rightarrow n\Lambda\Lambda$ and the 
${\bf \Lambda\Lambda}$ final state interaction}
\end{title}

\author{S.\ B.\ Carr and I.\ R.\ Afnan}
\address{Department of Physics, Faculty of Science and Engineering,
         Flinders University of South Australia,
         Bedford Park, SA 5042, Australia} 
\author{B.\ F.\ Gibson}
\address{Theoretical Division, Los Alamos National Laboratory, \\
         Los Alamos, NM 87545, USA}
\date{\today}

\maketitle

\begin{abstract}
The reaction $\Xi^-d\rightarrow n\Lambda\Lambda$ is studied within 
the framework of the Faddeev equations as a possible tool to gain insight
into the final state $\Lambda$--$\Lambda$ interaction. The neutron
differential energy spectrum gives a final state interaction that is
sensitive to both the $\Lambda$--$\Lambda$ amplitude at threshold, and the
coupling between the $\Lambda$--$\Lambda$ and $\Xi$--$N$ channels. The
latter is a result of interference between two mechanisms for the
production of the final state, which suggests that this reaction could
give a measure of flavor $SU(3)$ violation in the two-baryon system.
\end{abstract}


\section{Introduction}\label{sec.1}

Interest in the reaction $\Xi^-d\rightarrow n\Lambda\Lambda$ is twofold: 
(i)~It can be used to examine the final state interaction between the 
two $\Lambda$ hyperons and, therefore, to gain some insight into
the baryon-baryon interaction. The fact that this reaction requires
coupling between the $\Xi$--$N$ and $\Lambda$--$\Lambda$ channels to
proceed, suggests that one may gain further understanding of the 
importance of this coupling.  This reaction might possibly set a 
constraint on its magnitude, which in turn could give a measure of 
$SU(3)$ violation in this system as well as the general two baryon 
system.  (ii)~It can be used to test the hypothesis that there is an 
$H$ dibaryon, a six quark state. The existence of such a dibaryon 
would be a signature that meson-baryon degrees of freedom are not 
sufficient to describe this reaction and possibly the baryon-baryon
system in general. If the $H$ has a mass significantly below the 
$\Lambda$--$\Lambda$ threshold, then the present reaction would provide 
a clean signal (a mono energetic neutron) in the neutron differential 
energy spectrum (NDES). However, if the $H$ has a mass comparable to the 
$\Lambda$--$\Lambda$ threshold, then it would be difficult to distinguish, 
in the NDES, between the $H$ dibaryon and the final state 
$\Lambda$--$\Lambda$ interaction. In this case one needs to compare the 
experimental data with a theoretical calculation based upon meson-baryon 
dynamics. To that extent this investigation could be used as a guide in 
the search for $H$ dibaryons.

Aerts and Dover \cite{AD84} first examined the 
$\Xi^-d\rightarrow n\Lambda\Lambda$ reaction within the context of
estimating rates and spectra for the $(\Xi^-d)_{atom}\rightarrow n H$
reaction.  They approximated the decay rate for the
$(\Xi^-d)_{atom}\rightarrow n\Lambda\Lambda$ reaction by evaluating
the lowest order diagram, including no multiple scattering in the
$\Xi NN$ intermediate state or between the $\Lambda\Lambda$ pair in
the final state.  Taking into account such multiple scattering is 
essential if one is to model the enhancement in the neutron energy 
spectrum due to a strong $\Lambda\Lambda$ interaction; transfer of the 
kinetic energy from the $\Lambda\Lambda$ pair created by the 
$\Xi^-N\rightarrow\Lambda\Lambda$  transition to the neutron
requires a complex series of multiple scatterings in the final state
if the spectator neutron is to carry off almost all the available energy
and probe the $\Lambda\Lambda$ zero-energy scattering length.  This
is the analog of the $nd\rightarrow nnp$ breakup experiment (see, for 
example, Tornow {\it et al.}~\cite{TOR96} for a recent review) in which 
measurement of the proton energy spectrum was proposed as a means to 
study the $nn$ zero-energy scattering length.  The Watson-Migdal 
approximation fails to properly describe the proton energy spectrum;  
multiple scattering calculations \cite{AA66,CS71} which include fully 
three-body dynamics are essential. 

To examine the reaction $\Xi^-d\rightarrow n\Lambda\Lambda$ within 
the framework of three-body dynamics, we must derive a set of 
equations for this specific problem. The new feature, unique to this 
problem, is the fact that the initial $\Xi NN$ system and the final 
$\Lambda\Lambda N$ system have two (separate) identical spin 
$\frac{1}{2}$ particles, which must be in antisymmetric states. 
In Sec.~\ref{sec.2} we first derive the expression for the breakup
amplitude with both the initial two nucleons and the final two $\Lambda$
hyperons in antisymmetric states. We then proceed in Sec.~\ref{sec.2.2} 
to adapt the Alt-Grassberger-Sandhas (AGS) three-body equations 
\cite{AGS67} to generate the antisymmetric elastic and rearrangement
amplitudes required to determine the breakup amplitude. In this way
we minimize the number of coupled integral equations needed to extract
the final breakup amplitude. In Sec.~\ref{sec.2.3} we write 
the breakup cross section as the coherent sum of three reaction
mechanisms. To calculate the NDES for this reaction, we define the
input two-body amplitudes in Sec.~\ref{sec.3.1}, and demonstrate in
Sec.~\ref{sec.3.2} that the final amplitudes have been accurately
calculated to satisfy three-body unitarity. The fact that the final
state interaction between the two $\Lambda$ hyperons is part of the
coupled channel $\Lambda\Lambda$--$\Xi N$ problem implies that there
are two ways of converting the $\Xi$ to a $\Lambda$, the first is in
the final state interaction [see diagram (a) of Fig.~\ref{fig.1}], the
second in the rearrangement amplitude [as in diagram (b) of
Fig.~\ref{fig.1}]. We find in Sec~\ref{sec.3.3} that these two
amplitudes, each dominated by the final interaction in the 
$\Lambda\Lambda$--$\Xi N$ channel, are out of phase and almost cancel 
one another. It is only when the background term from diagram (c) 
of Fig.~\ref{fig.1} is also included in the evaluation of the cross 
section that a relatively weak final state interaction peak is present
in the NDES, and only if the $\Lambda\Lambda$--$\Xi N$ amplitude is
dominated at low energies by a virtual bound state pole near the
$\Lambda$--$\Lambda$ threshold. The cancellation between diagrams
(a) and (b) of Fig.~\ref{fig.1} results in a final state interaction
peak that is sensitive to the coupling between the
$\Lambda$--$\Lambda$ and the $\Xi$--$N$ channels. Finally, in
Sec.~\ref{sec.4} we present some concluding remarks about the
implication of the results.


\section{Theory}\label{sec.2}

The reaction $\Xi^- d\rightarrow n\Lambda\Lambda$ differs from the 
standard three-body problem, {\it e.g.} $nd\rightarrow nnp$, in that 
the particles in the initial state $\Xi NN$ differ from those in the 
final state $N\Lambda\Lambda$, and in that both initial and final 
states contain two different pairs of identical Fermions, for which 
we need to insure antisymmetric wave functions. 
In addition, the $\Xi N\rightarrow \Lambda\Lambda$ conversion can take 
place on either nucleon in the initial state. All of these features 
can be included naturally if we work within the framework of $SU(3)$ 
rather than the $SU(2)$ of isospin. However, then we cannot include the 
mass splitting in the baryon octet, and we must carry out the 
necessary re-coupling within the framework of $SU(3)$ algebra. 
To avoid this complication, and to take into consideration the mass 
splitting in the baryon octet, we have resorted to antisymmetrizing 
explicitly both initial and final states. In this section 
we first derive an expression for the antisymmetric breakup
amplitude describing the reaction $\Xi^- d\rightarrow n\Lambda\Lambda$ 
in terms of antisymmetric off-shell elastic and rearrangement
amplitudes. We then proceed, with the help of the AGS equations to 
derive coupled integral equations for the elastic and rearrangement
amplitudes. By deriving coupled integral equations for the 
antisymmetric amplitude, we reduce the number of coupled integral 
equations we need to solve. Finally, we write the cross 
section for this reaction in terms of these antisymmetric amplitudes.

\subsection{Antisymmetry}\label{sec.2.1} 

As a first step in determining the antisymmetric amplitude for 
$\Xi d\rightarrow N\Lambda\Lambda$, we construct the initial
and final states for this reaction. For the initial $\Xi d$ state, we 
designate the two nucleons as particles 1 and 2 and the $\Xi$ as 
particle 3, while for the final state the nucleon can be either 
particle 1 or 2 depending on which nucleon the $\Xi$ converted into 
a $\Lambda$. We now introduce the antisymmetrization operator 
${\cal A}_{ij}$ defined in terms of the permutation operator  
$P_{ij}$ that exchanges the coordinates of particles $i$ and $j$ 
as~\cite{GW64}:
\begin{equation}
{\cal A}_{ij} = \frac{1}{2} \left( 1 - P_{ij} \right)   \ ,   \label{eq:1}
\end{equation}
where ${\cal A}_{ij}{\cal A}_{ij}={\cal A}_{ij}$.
This allows us to define an initial state that is antisymmetric in the two
nucleons that form the deuteron as
\begin{equation}
|\Xi^-\,d\rangle^{AS} = {\cal A}_{12}\,|\Xi^-\,d\rangle\ ,  \label{eq:2}
\end{equation}
with the antisymmetrized state being normalized; {\it i.e.},
\begin{equation}
^{AS}\langle d\,\Xi^-|\Xi^-\,d\rangle^{AS} = 1\ .    \label{eq:3}
\end{equation}
We note that, if the deuteron is in the $^3$S$_1$-$^3$D$_1$ channel, then
${\cal A}_{12}\,|\Xi^-\,d\rangle=|\Xi^-\,d\rangle$.

For the final state we have the two $\Lambda$ hyperons which must be in 
an antisymmetric state. However, now we need to recall that the final nucleon
can be either nucleon 1 or nucleon 2, depending on which nucleon the $\Xi$ 
converted to a $\Lambda$. We therefore have two possible 
configurations\cite{foot1}:
\begin{eqnarray}
|N_1\Lambda\Lambda\rangle^{AS} &=& \sqrt{2}{\cal A}_{23}
                            |N_1\Lambda_2\Lambda_3\rangle \nonumber \\
|\Lambda N_2\Lambda\rangle^{AS} &=& \sqrt{2}{\cal A}_{13}
                            |\Lambda_1 N_2\Lambda_3\rangle\ . \label{eq:4}
\end{eqnarray}
The factor of $\sqrt{2}$ was introduced to guarantee that the antisymmetric 
states are normalized to one. Since neither of these states is the physical 
state, we need to take the linear combination of these states such that the
final state is antisymmetric with respect to $N_1$ and $N_2$, {\it i.e.}
\begin{eqnarray}
|N\Lambda\Lambda\rangle^{AS} &=& \sqrt{2}{\cal A}_{12}
                          |N_1\Lambda\Lambda\rangle^{AS} \nonumber\\
   &=&{\cal A}_{23}|N_1\Lambda_2\Lambda_3\rangle 
    - {\cal A}_{13}|\Lambda_1 N_2\Lambda_3\rangle\ .    \label{eq:5} 
\end{eqnarray}
In writing the above result, we have made use of the multiplication table 
of the permutation operators, and in particular the fact that
\begin{equation}
P_{12}\,P_{23} = P_{123} = P_{13}\,P_{12}\ .              \label{eq:6}
\end{equation}
We will see next that the antisymmetry in the initial state between 
the two nucleons propagates through a symmetric breakup operator to 
project only the antisymmetric combination considered in 
Eq.~(\ref{eq:5}).

We are now in a position to define the physical amplitude for 
$\Xi d\rightarrow N\Lambda\Lambda$ as the matrix element of the breakup
operator for distinguishable particles between antisymmetric and normalized
initial and final states~\cite{AT77}; {\it i.e},
\begin{eqnarray}
T_{N\Lambda\Lambda\leftarrow\Xi d} &=& ^{AS}\langle N\Lambda\Lambda|\,
                                    U_{03}|\Xi d\rangle^{AS}\nonumber \\
&=& ^{AS}\langle N\Lambda\Lambda|\,\sum_\alpha\,T_\alpha G_0 U_{\alpha 3}\,
   |\Xi d\rangle^{AS}\ .                             \label{eq:7}
\end{eqnarray}
We note here that $U_{03}$ commutes with the permutation operator $P_{12}$,
and it is this feature of the breakup operator that required we write the
antisymmetrized final state in the form given in Eq.~(\ref{eq:5}).
In Eq.~(\ref{eq:7}), the operators $U_{\alpha 3}$ are the AGS~\cite{AGS67} 
operators that are a solution of the AGS equations
\begin{equation}
U_{\alpha \beta}(E) = \bar{\delta}_{\alpha \beta} G_{0}^{-1}(E) 
                    + \sum_{\gamma} \bar{\delta}_{\alpha \gamma} T_{\gamma}(E) 
                    G_{0}(E) U_{\gamma \beta}(E)\ .     \label{eq:8}
\end{equation}
Here $\bar{\delta}_{\alpha\beta}=1-\delta_{\alpha\beta}$, while $G_0$ is 
the free three-body Green's function, and $T_\alpha$ the amplitude for 
particles $\beta$ and $\gamma$.

We are now in a position to write the physical amplitude for
$\Xi d\rightarrow N\Lambda\Lambda$ in terms of a linear combination of 
the AGS operators $U_{\alpha\beta}$ to maintain the antisymmetry in 
both initial and final states. We then can use the 
AGS equations to derive integral equations for these antisymmetric
combinations of $U_{\alpha\beta}$. The advantage of this procedure is 
a reduction in the number of coupled integral equations we need to solve 
to construct the breakup amplitude. 

Combining Eq.~(\ref{eq:5}) with Eq.~(\ref{eq:7}), we can write the
physical amplitude as:
\begin{eqnarray}
 T_{N\Lambda\Lambda\leftarrow\Xi d} & = & \langle N_1\Lambda_2\Lambda_3| 
{\cal A}_{23} \sum_{\alpha} T_{\alpha} G_{0} U_{\alpha 3} 
{\cal A}_{12}|\Xi d \rangle \nonumber \\
 & - & \langle\Lambda_1N_2\Lambda_3| {\cal A}_{13}\sum_{\alpha}T_{\alpha}
G_{0}U_{\alpha 3}{\cal A}_{12}|\Xi d\rangle \ .    \label{eq:9}
\end{eqnarray}
The permutation operator $P_{12}$ exchanges the coordinate of particles 1 
and 2. In the final state, this interchanges the position of the particle in
the ket and the label on the nucleon; {\it i.e.},
\begin{equation}
\langle\Lambda_1N_2\Lambda_3| = \langle N_1\Lambda_2\Lambda_3|
                                \,P_{12}\ .        \label{eq:10}
\end{equation}
With the help of the identities
\begin{equation}
P_{12}\,{\cal A}_{13} = {\cal A}_{23}\,P_{12}\ \,\quad
P_{12}{\cal A}_{12} = - {\cal A}_{12}             \label{eq:11}
\end{equation}
that follow from the multiplication table of the permutation operators,
we can write the breakup amplitude as
\begin{eqnarray}
 T_{N\Lambda\Lambda\leftarrow\Xi d} 
     & = & 2\,\langle N_1\Lambda_2\Lambda_3|\,{\cal A}_{23}\,{\cal A}_{12} 
           \, U_{0 3}\,{\cal A}_{12}|\Xi d \rangle    \nonumber \\
     & = & 2\, \langle N_1\Lambda_2\Lambda_3|\,{\cal A}_{23}
           \,{\cal A}_{12}\,\sum_{\alpha}\ T_{\alpha} G_{0} U_{\alpha 3} 
          {\cal A}_{12}|\Xi d \rangle        \ .  \label{eq:12}
\end{eqnarray}
In writing the above expression for the breakup amplitude, we have 
maintained the antisymmetry operators on both sides of the breakup 
amplitude $U_{03}$.
To reduce the right hand side of the above expression for the 
breakup amplitude in terms of antisymmetric two-cluster final state 
amplitudes, we make use of the fact that
\begin{equation}
T_\alpha\,P_{\alpha\beta}=P_{\alpha\beta}\,T_\beta\ ,\quad\mbox{and}\qquad
T_\gamma\,P_{\alpha\beta}=P_{\alpha\beta}\,T_\gamma \quad \mbox{for}\quad
\alpha\neq\gamma\neq\beta\ ,                          \label{eq:13}
\end{equation}
and operate with ${\cal A}_{12}$ in the final state on the breakup 
amplitude. This will assure us that the final state is antisymmetric 
in particles 1 and 2. The resultant breakup amplitude is the sum of 
two terms. The first has a final state interaction (FSI) between the two 
$\Lambda$ hyperons, while the second term has an $N\Lambda$ FSI; 
{\it i.e.},
\begin{eqnarray}
T_{N\Lambda\Lambda\leftarrow\Xi d} 
  & = & \,\langle N_1\Lambda_2\Lambda_3|\,{\cal A}_{23}\,T_1\,G_0\,
         \left[\,U_{13} - P_{12}\,U_{23}\,\right]\,
        {\cal A}_{12}|\Xi d \rangle               \nonumber \\
  &   &\ +  \langle N_1\Lambda_2\Lambda_3|{\cal A}_{23}\,T_2\,G_0\,
       \left[\,(U_{23} - P_{23}\,U_{33}) - P_{12}\,(U_{13} - 
       P_{13}\,U_{33})\,\right]\,{\cal A}_{12}\,|\Xi d \rangle 
       \ .                                          \label{eq:14}
\end{eqnarray}
The linear combination of the amplitudes in Eq.~(\ref{eq:14}) insures
that the antisymmetry in the final state is preserved at the operator 
level.

The AGS equations take the form of a closed set of coupled integral 
equations for physical amplitudes only if the two-body interaction is 
assumed to be separable. Since we will be using separable interactions for 
calculating the cross section for this reaction, it is most convenient 
to introduce this approximation at this time as it will give a physical 
meaning to the terms resulting from Eq.~(\ref{eq:14}). Production of the 
$N\Lambda\Lambda$ final state will require the conversion of $\Xi N$ 
into $\Lambda\Lambda$. This can take place either in the two-body amplitude 
$T_1$ in the FSI where the conversion takes place 
on nucleon 2, or in the three-body AGS amplitudes 
$\left[\,U_{13}-P_{12}\,U_{23}\,\right]$. To expose the mechanism 
for conversion, we need to write the amplitude $T_1$ in the form of 
$2\times 2$ matrix. In three-body Hilbert space this takes the form
\begin{equation}
T_{1}(E) = \sum_{\alpha\beta}\ |g^{c}_{\alpha};N_{1}\rangle \ 
           \tau^{c}_{\alpha\beta}(E-\epsilon_{1})\ 
           \langle g^{c}_{\beta};N_{1}|    \ ,      \label{eq:15}
\end{equation}
where the sum $\alpha,\beta$ runs over the coupled channels 
$N\Xi, \Lambda\Lambda$, and the superscript $^{c}$ indicates that 
this is a coupled channel partial wave. Since we have written the 
two-body amplitude $T_{1}(E)$ in three-body Hilbert space, 
$\epsilon_{1}$ is the energy of the spectator nucleon 1.

Because the $\Lambda\Lambda$ is an isospin zero system, this matrix 
structure for $T_{1}(E)$ is only present for this isospin channel and 
partial waves in which the space-spin wave function are antisymmetric. 
In all other partial waves the $\Xi N$ system does not couple to the 
$\Lambda\Lambda$ channel, and the corresponding amplitude is a single 
channel amplitude, and has the same form as the $N\Lambda$ interaction, 
which can be written as
\begin{equation}
T_i(E) = |g_{N\Lambda_j};N_{i}\rangle\ \tau_{N\Lambda}(E-\epsilon_{i})\ 
         \langle g_{N\Lambda_j}:N_{i}|
\quad j\neq i=2,3 .                                    \label{eq:16} 
\end{equation}
In writing the separable representation for the $N\Lambda$ amplitude, 
we have excluded the particle label from the $\tau_{N\Lambda}(E)$ because 
this quantity is the same for the nucleon interacting with either 
$\Lambda$.

Making use of the above separable representation for the two-body amplitude, 
we can write the amplitude for $N\Lambda\Lambda\leftarrow\Xi d$ as
\begin{eqnarray}
T_{N\Lambda\Lambda\leftarrow\Xi N}
    &=& \langle N_{1}\Lambda_2\Lambda_3|\,{\cal A}_{23}\,
        |g^{c}_{\Lambda\Lambda};N_{1}\rangle\,
        \left[\,\tau^{c}_{\Lambda\Lambda;\Xi N}\,X^{c,}_{N;\Xi}
        + \tau^{c}_{\Lambda\Lambda;\Lambda\Lambda}\,Y_{N;\Xi}\,
           \right]                                     \nonumber\\
    & &\ +\ 2\ \langle N_1\Lambda_{2}\Lambda_3|\,{\cal A}_{23}\,
          |g_{N\Lambda};N_{2}\rangle\ \tau_{N\Lambda}
           \,Y_{\Lambda;\Xi}          \ .         \label{eq:17}
\end{eqnarray}
The first two terms on the right hand side correspond to the final 
interaction occuring in the $N\Xi$--$\Lambda\Lambda$ channel, while the 
last term has the interaction in the $N\Lambda$ channel. The 
antisymmetric AGS amplitudes $X_{\alpha\beta}$ and $Y_{\alpha\beta}$ 
are defined as
\begin{eqnarray}
X_{N;\Xi} &\equiv& \frac{1}{\sqrt{2}}\ \langle g_{\Xi N};N_{1}|\,G_{0}\,
               \left[\,U_{13} - P_{12}\,U_{23}\,\right]\,G_{0}\,
              |g_{NN};\Xi\rangle                       \nonumber \\
          &=& \frac{1}{\sqrt{2}}\ \left[X_{N_{1};\Xi} - 
               X_{N_{2};\Xi}\,\right]             \label{eq:18}
\end{eqnarray}
 for the antisymmetric $(\Xi N)N\leftarrow\Xi d$, while the antisymmetric 
amplitude for $(\Lambda\Lambda)N\leftarrow\Xi d$ is defined as
\begin{eqnarray}
Y_{N;\Xi} &\equiv& \frac{1}{\sqrt{2}}\ \langle g_{\Lambda\Lambda};N_{1}|\,
              \,G_{0}\,\left[\,U_{13} - P_{12}\,U_{23}\,\right]
               \,G_{0}\,|g_{NN};\Xi\rangle        \nonumber \\
          &=& \frac{1}{\sqrt{2}}\ \left[Y_{N_{1};\Xi} - 
               Y_{N_{2};\Xi}\,\right]   \ .      \label{eq:19}
\end{eqnarray}
In writing the above two definitions for the amplitudes, we have 
dropped the superscript as the definitions are valid for both coupled 
and uncoupled channels. Finally the antisymmetric amplitude for the 
reaction $(N\Lambda)\Lambda\leftarrow\Xi d$ is given by
\begin{eqnarray}
Y_{\Lambda;\Xi} &\equiv& \frac{1}{2}\ \langle 
                   g_{N_{1}\Lambda_{3}};\Lambda_{2}|\,G_{0}\left[\,
                   (U_{23} - P_{23}\,U_{33}) 
                   - P_{12}\,(U_{13} - P_{13}\,U_{33})\right]\,
                   G_{0}\,|g_{NN};\Xi\rangle     \nonumber \\
                &=& \frac{1}{2}\ \left[\,Y^{N_{1}}_{\Lambda_{2};\Xi}
                    - Y^{N_{1}}_{\Lambda_{3};\Xi}
                    - Y^{N_{2}}_{\Lambda_{1};\Xi}
                    + Y^{N_{2}}_{\Lambda_{3};\Xi}
                    \right]\ .                     \label{eq:20}
\end{eqnarray}
In defining $X_{\alpha\beta}$ and $Y_{\alpha\beta}$, we have made use 
of the fact that in a separable approximation we can write the initial 
state $|\Xi d\rangle = G_0\,|g_d;\Xi\rangle$. In addition, it is assumed 
that the deuteron is in an antisymmetric state; {\it i.e.}, 
${\cal A}_{12}|g_{d};\Xi\rangle =  |g_{d};\Xi\rangle$.
The amplitude given in Eq.~(\ref{eq:17}) can be represented 
diagrammatically as in Fig.~\ref{fig.1}.

\subsection{The AGS equations for $\Xi d\rightarrow N\Lambda\Lambda$}
\label{sec.2.2}

We are now in a position to derive integral equations for the 
antisymmetric amplitudes $X_{\alpha\beta}$ and $Y_{\alpha\beta}$ defined 
in Eqs.~(\ref{eq:18})-(\ref{eq:20}), and required in  Eq.~(\ref{eq:17}) to 
construct the total amplitude for breakup. In addition to these amplitudes, 
we need the amplitude for $\Xi d$ elastic scattering, which is basically 
a matrix element of $U_{33}$. Thus for $X_{N;\Xi}$ we have, after making use 
of the AGS equation,
\begin{eqnarray}
X^{c,}_{N;\Xi} &\equiv& \frac{1}{\sqrt{2}}\ \langle g^{c}_{\Xi N};N_{1}|\,
              G_{0}\,\left[\,U_{13} - P_{12}\,U_{23}\,\right]\,G_{0}\,
              |g_{NN};\Xi\rangle                       \nonumber \\
          &=& \frac{1}{\sqrt{2}}\ \langle g^{c}_{\Xi_{3}N_{2}};N_1|\,
              (1-P_{12})\,G_0|g_{NN};\Xi\rangle        \nonumber \\
      & & \  + \frac{1}{\sqrt{2}}\ \langle g^{c}_{\Xi_{3}N_{2}};N_1|\,G_0
               \,T_2\,G_0\,\left(\,U_{23} - P_{12}\,U_{13}\,\right)\,
               G_0\,|g_{NN};\Xi\rangle                \nonumber \\
      & & \  + \frac{1}{\sqrt{2}}\ \langle g^{c}_{\Xi_{3}N_{2}};N_1|\,
               (1-P_{12})\,G_0\,T_{3}\,G_{0}\,U_{33}\,G_0\,
               |g_{NN};\Xi\rangle\ .                  \label{eq:21}
\end{eqnarray}
Introducing the separable representation for the amplitudes $T_{2}$ and 
$T_{3}$ allows us to turn the above expression into an integral 
equation of the form
\begin{eqnarray}
X^{c,}_{N;\Xi} &=& Z^{c,}_{N;\Xi} + Z^{c,}_{N;N}\ \tau_{\Xi N}\ X_{N;\Xi}
          + Z^{c,c}_{N;N}\ \tau^{c}_{\Xi N;\Xi N}\ X^{c,}_{N;\Xi}\nonumber\\
     & &\ + Z^{c,c}_{N;N}\ \tau^{c}_{\Xi N;\Lambda\Lambda}\ Y^{c,}_{N;\Xi}
          + Z^{c,}_{N;\Xi}\ \tau_{NN}\ X_{\Xi;\Xi}\ ,\label{eq:22}     
\end{eqnarray}
where the antisymmetric elastic amplitude is given by
\begin{equation}
X_{\Xi;\Xi} \equiv\langle g_{NN};\Xi|\,G_{0}\,U_{33}\,G_{0}\,
              |g_{NN};\Xi\rangle       \ ,           \label{eq:23}
\end{equation}
while the antisymmetric Born amplitudes $Z_{\alpha\beta}$ needed in 
Eq.~(\ref{eq:22}) are defined 
as
\begin{eqnarray}
Z_{N;\Xi} &\equiv& \frac{1}{\sqrt{2}}\ \langle g_{\Xi_{3}N_{2}}:N_{1}|\,
              (1-P_{12})\,G_{0}\,|g_{NN};\Xi\rangle  \nonumber \\
          &=& \frac{1}{\sqrt{2}}\ \left[\,Z_{N_{1};\Xi} - 
               Z_{N_{2};\Xi}\,\right]                 \label{eq:24}
\end{eqnarray}
and
\begin{eqnarray}
Z_{N;N} &\equiv& - \langle g_{\Xi_{3}N_{2}};N_{1}|\,G_{0}\,
                   | g_{\Xi N};N_{2}\rangle              \nonumber \\
        &=& -\ Z_{N_{1};N_{2}}     \ .             \label{eq:25}
\end{eqnarray}
In Eq.~(\ref{eq:22}) we have the amplitude $X^{c,}_{N;\Xi}$ in terms of
the amplitudes $X^{c,}_{N;\Xi}$, $Y^{c,}_{N;\Xi}$, $X_{N;\Xi}$, and 
$X_{\Xi;\Xi}$. To close this set of integral equations, we generate 
an equation for each of these amplitudes following the same 
procedure we adopted for deriving Eq.~(\ref{eq:22}). For the amplitude
$X_{N;\Xi}$, we have by definition the same expression as in
Eq.~(\ref{eq:21}) without the superscript $^c$. This results in an
equation for the $X_{N;\Xi}$ amplitude which is identical to
Eq.~(\ref{eq:22}) without the left hand superscript $^c$.

For the amplitude $Y^{c,}_{N;\Xi}$, we have
\begin{eqnarray}
Y^{c,}_{N;\Xi} &\equiv& \frac{1}{\sqrt{2}}\ 
               \langle g^{c,}_{\Lambda\Lambda};N_{1}|\,G_{0}
               \,\left[\,U_{13} - P_{12}\,U_{23}\,\right]\,G_{0}\,
               |g_{NN};\Xi\rangle                    \nonumber \\
           &=& \frac{1}{\sqrt{2}}\ 
               \langle g^{c,}_{\Lambda\Lambda};N_{1}|\,G_{0}\,T_{2}\,
               G_{0}\,\left[\,U_{23} - P_{12}\,U_{13}\,\right]\,G_{0}\,
               |g_{NN};\Xi\rangle                    \nonumber \\
         &&\ +\frac{1}{\sqrt{2}}\ 
               \langle g^{c,}_{\Lambda\Lambda};N_{1}|\,G_{0}\,T_{3}\,
               G_{0}\,\left[\,U_{33} - P_{12}\,U_{33}\,\right]\,G_{0}\,
               |g_{NN};\Xi\rangle            \ .     \label{eq:26}
\end{eqnarray}
Making use of the fact that $T_{3}=P_{23}\,T_{2}\,P_{23}$ and 
$\langle g^{c}_{\Lambda\Lambda};N_{1}|P_{23} = 
-\langle g^{c}_{\Lambda\Lambda};N_{1}|$, the integral equation for 
$Y_{N;\Xi}$ reduces to
\begin{equation}
Y^{c,}_{N;\Xi} = Z^{c,}_{N;\Lambda}\ \tau_{N\Lambda}\ 
                 Y_{\Lambda;\Xi}\ ,                    \label{eq:27}
\end{equation}
where
\begin{eqnarray}
Z_{N;\Lambda} &\equiv& \sqrt{2}\ \langle g^{c,}_{\Lambda\Lambda};N_{1}|
                       \,G_{0}\,|g_{N\Lambda};\Lambda\rangle\nonumber\\
              &=& \sqrt{2}\ Z^{c,}_{N_{1};\Lambda_{2}}\ .\label{eq:28}
\end{eqnarray}
In this same way we establish an equation for $Y_{\Lambda;\Xi}$ to be
\begin{eqnarray}
Y_{\Lambda;\Xi} &=& Z^{\ ,c}_{\Lambda;N}\ 
               \tau^{c}_{\Lambda\Lambda;\Xi N}\ X^{c,}_{N;\Xi}
             + Z^{\ ,c}_{\Lambda;N}\ 
               \tau^{c}_{\Lambda\Lambda;\Lambda\Lambda}\ 
               Y^{c,}_{N;\Xi}                     \nonumber \\
           & &\ + Z_{\Lambda;\Lambda}\ \tau_{N\Lambda}\ 
                  Y_{\Lambda;\Xi}       \ ,           \label{eq:29}
\end{eqnarray}
where $Z_{\Lambda;N}$ is defined as
\begin{eqnarray}
Z_{\Lambda;N} &\equiv& \frac{1}{\sqrt{2}}\ 
              \langle g_{N_{1}\Lambda_{3}};\Lambda_{2}|\,(1-P_{23})\,
              G_{0}\,|g_{\Lambda\Lambda};N_{1}\rangle \nonumber \\
              &=& \sqrt{2}\ Z_{\Lambda_{2};N_{1}}\ ,  \label{eq:30}
\end{eqnarray}
and
\begin{equation}
Z_{\Lambda;\Lambda} = -\ Z_{\Lambda_{3};\Lambda_{2}}\ .\label{eq:31}
\end{equation}
Finally to close the set of coupled equations, we must write the 
equation for the elastic amplitude. Making use of the definition of 
$X_{\Xi;\Xi}$, given in Eq.~(\ref{eq:23}), we get
\begin{eqnarray}
X_{\Xi;\Xi} &=& Z_{\Xi;N}\ \tau_{\Xi N}\ X_{N;\Xi} +
                Z^{\ ,c}_{\Xi;N}\ \tau^{c}_{\Xi N;\Xi N}\ 
                X^{c,}_{N;\Xi}                        \nonumber \\
            &&\ + Z^{\ ,c}_{\Xi;N}\ \tau^{c}_{\Xi N;\Lambda\Lambda}\ 
                  Y^{c,}_{N;\Xi} \ ,                     \label{eq:32}
\end{eqnarray}
where
\begin{eqnarray}
Z_{\Xi;N} &\equiv& \frac{1}{\sqrt{2}}\ 
              \langle g_{NN};\Xi|\,(1-P_{12})\,G_{0}\,
              |g_{\Xi N};N\rangle                     \nonumber \\
          &=& \frac{1}{\sqrt{2}}\ 
               \left[\,Z_{\Xi_{3};N_{1}} - Z_{\Xi_{3};N_{2}}\,\right]
                     \ .                              \label{eq:33}
\end{eqnarray}
Equations (\ref{eq:22}), (\ref{eq:27}), (\ref{eq:29}), and 
(\ref{eq:33}) now form a closed set for the two-cluster to two-cluster 
antisymmetrized amplitudes required to construct the breakup amplitude 
and, therefore, the cross section. In the appendix we provide a summary 
of the antisymmetrized Born amplitudes in terms of 
the single particle exchanged amplitudes.

\subsection{The cross section for $\Xi d\rightarrow N\Lambda\Lambda$}
\label{sec.2.3}

We now turn to the determination of the cross section, and in 
particular the neutron energy spectrum for the capture of $\Xi^{-}$ 
on the deuteron in terms of the amplitudes defined in 
Sec.~\ref{sec.2.1}. Since our ultimate aim is to gain some insight 
into the low energy $\Lambda\Lambda$ interaction in a reaction in 
which the $\Xi^{-}$ is captured from an atomic orbit, we will 
restrict our analysis to $S$-wave two-body interactions. With this 
simplification, the only angular momentum in the problem is the spin 
of the three particles, and the total spin will be either $\frac{1}{2}$ 
(doublet) or $\frac{3}{2}$ (quartet). However, because the two $\Lambda$ 
hyperons are in a $^{1}$S$_{0}$ state, only the doublet channel 
contributes. In addition, by including isospin and taking into 
consideration the fact that the two $\Lambda$ hyperons and the deuteron 
are in an isospin zero state, the total isospin is $\frac{1}{2}$ with a 
projection $-\frac{1}{2}$, corresponding to a $\Xi^{-}$ in the initial 
state. With these restrictions, we can write the breakup amplitude as
the sum of the three diagrams in Fig.~\ref{fig.1} in the spin 
isospin $\frac{1}{2}$ state as:
\begin{eqnarray}
T_{\Lambda \Lambda n \leftarrow\Xi^{-} d} & = & 
\sum_{S_{N}}\sum_{{\cal M_{S}} = \pm \frac{1}{2}}\   
({\textstyle\frac{1}{2}}\,m_{s_{2}}\,{\textstyle\frac{1}{2}}\,m_{s_{3}} 
                         | S_{N} M_{S_{N}} ) 
(S_{N}\, M_{S_{N}}\, {\textstyle\frac{1}{2}}\, m_{s_{1}} 
                |{\textstyle \frac{1}{2}}\, {\cal M_{S}} ) 
                                                     \nonumber \\ 
    &&\ \times\ \left[\, A_{\ell_{N}:\ell_{\Xi}} 
                       + B_{\ell_{N};\ell_{\Xi}}
                       + C_{\ell_{N};\ell_{\Xi}}\,\right]\   
    (S_{d}\, M_{S_{d}}\, {\textstyle\frac{1}{2}}\, m_{s_{\Xi}}|
    {{\textstyle \frac{1}{2}}\, M_{S}}) 
                 \ ,                                 \label{eq:34}  
\end{eqnarray} 
where $\ell_{\alpha}=\{S_{\alpha},t_{\alpha}\}$, with $S_{\alpha}$ and 
$t_{\alpha}$ the spin and isospin of the pair $(\beta\gamma)$ 
respectively. This basically defines the channels in the two-body 
subsystems. The amplitude $A_{\ell_{N};\ell_{\Xi}}$ corresponding to 
Fig.~\ref{fig.1}(a) is given in terms of the AGS amplitude 
$X_{\alpha\beta}$ by
\begin{eqnarray}
A_{\ell_{N};\ell_{\Xi}}({\bf p}_{1},{\bf k}_{1};{\bf k}_{\Xi}^{0})
  &=& \frac{\delta_{S_{N},0}}{\sqrt{ 2\pi}}\ g_{\ell_{N}}(p_{1})\ 
      \tau_{\Lambda\Lambda;\Xi N}[ E - \epsilon (k_{1})] \ 
      \ X_{\ell_{N};\ell_{\Xi}}({\bf k}_{1}, {\bf k}_{\Xi}^{0})
                                     \ ,           \label{eq:35}
\end{eqnarray}
where ${\bf k}_{\Xi}^{0}$ and ${\bf k}_{1}$ are the initial on-shell 
momentum of the $\Xi^{-}$ and the final momentum of the neutron 
respectively. The momentum $p_{1}$ in Eq.~(\ref{eq:35}) is the 
relative momentum of the two $\Lambda$ hyperons in the final state. 
The AGS amplitude $X_{\ell_{N};\ell_{\Xi}}$ is that 
defined in Eq.~(\ref{eq:18}) for a state with total spin and isospin 
$\frac{1}{2}$. Here we note that the $\delta_{S_{N}0}$ restricts the 
contribution of Fig.~\ref{fig.1}(a) to the amplitude in which the 
$\Lambda\Lambda$ interaction occurs in the $^{1}$S$_{0}$ state. The 
amplitude corresponding to diagram Fig.~\ref{fig.1}(b), in which the 
conversion from $\Xi N$ to $\Lambda\Lambda$ takes place in the 
multiple scattering before the FSI, is given by
\begin{eqnarray}
B_{\ell_{N};\ell_{\Xi}}({\bf p}_{1},{\bf k}_{1};{\bf k}_{\Xi}^{0})
  &=& \frac{\delta_{S_{N},0}}{\sqrt{ 2\pi}}\ g_{\ell_{N}}(p_{1})\ 
      \tau_{\Lambda\Lambda;\Lambda\Lambda}[ E - \epsilon (k_{1})] \ 
      \ Y_{\ell_{N};\ell_{\Xi}}({\bf k}_{1}, {\bf k}_{\Xi}^{0})
                                     \ ,           \label{eq:36}
\end{eqnarray}
where the AGS amplitude $Y_{\ell_{N};\ell_{\Xi}}$ is defined in 
Eq.~(\ref{eq:19}) and is non-zero in total spin, isospin $\frac{1}{2}$ 
with the two $\Lambda$ hyperons in the $^{1}$S$_{0}$ state. If we compare 
the full breakup amplitude given in Eq.~(\ref{eq:17}) and (\ref{eq:34}), 
we observe that the antisymmetrization operator ${\cal A}_{23}$ in 
the first two terms on the right hand side of Eq.~(\ref{eq:17}) are not 
present in the expressions for $A_{\ell_{N};\ell_{\Xi}}$ and 
$B_{\ell_{N};\ell_{\Xi}}$. This is due to the fact that the 
$\Lambda\Lambda$ interaction in the final state has been restricted 
to those channels that satisfy the Pauli principle for the $\Lambda$ hyperons. 
This was simple to achieve by proper choice of partial waves. 
Unfortunately, that is not possible for the last term in 
Eq.~(\ref{eq:17}), and in this case we need to include the 
antisymmetrization operator explicitly. In addition, we need to 
maintain the same spin coupling as the diagrams in 
Figs.~\ref{fig.1}(a) and \ref{fig.1}(b). This will involve a 
re-coupling of the spin, with the resultant amplitude given by:
\begin{eqnarray}
C_{\ell_{N};\ell_{\Xi}}({\bf p}_{1},{\bf k}_{1};{\bf k}_{\Xi}^{0})
  &=& \sum_{S_{\Lambda}}\ (-1)^{R}\ \hat{S}_{N}\ \hat{S}_{\Lambda}\  
         \left\{ 
         \begin{array}{ccc}
        \frac{1}{2} & \frac{1}{2} & S_{N} \\
        \frac{1}{2} & \frac{1}{2} & S_{\Lambda}
        \end{array} \right\}                       \nonumber \\
   &&\quad \times \left[\,C^{(2)}_{\ell_{\Lambda};\ell_{\Xi}}
                         ({\bf p}_{2},{\bf k}_{2};{\bf k}_{\Xi}^{0}) 
        +  (-1)^{R'}\ C^{(3)}_{\ell_{\Lambda};\ell_{\Xi}}
                       ({\bf p}_{3},{\bf k}_{3};{\bf k}_{\Xi}^{0})\,
        \right]                                    \label{eq:37}
\end{eqnarray}
where $\hat{a}=\sqrt{2a+1}$, $R = S_{\Lambda}$ and, 
$R'=1-S_{N}$. The phase $(-1)^{R'}$ ensures that the two 
$\Lambda$ hyperons are in states that satisfy the Pauli principle. The 
amplitude $C^{(i)}_{\ell_{\Lambda};\ell_{\Xi}},\ i=2,3$ is given by
\begin{equation}
C^{(i)}_{\ell_{\Lambda};\ell_{\Xi}}
   ({\bf p}_{i},{\bf k}_{i};{\bf k}_{\Xi}^{0}) 
   = \frac{1}{\sqrt{4 \pi}}\, g_{\ell_{\Lambda}}(p_{i})\ 
     \tau_{\Lambda N}[ E - \epsilon (k_{i})] \ 
     Y_{\ell_{\Lambda};\ell_{\Xi}}(\vec{k}_{i}, \vec{k}_{\Xi}^{0}) 
         \ ,                                       \label{eq:38}
\end{equation}
where the AGS amplitude $Y_{\ell_{\Lambda};\ell_{\Xi}}$ is defined in 
Eq.~(\ref{eq:20}).

To calculate the neutron differential energy spectrum (NDES) we  
square the breakup amplitude, perform spin averaging over the 
incoming particles, and integrate over all the kinematical allowed final 
states for the two $\Lambda$ hyperons. This gives the 
cross section as a function of the final neutron energy. In the three-body 
center of mass system, the NDES takes the form
\begin{eqnarray}
\frac{d^{3} \sigma}{d E_{n} d \Omega_{n}} 
     & = &\ \frac{(2\pi)^{4}\mu_{\Xi d}k_{1}m_{N}}{2k_{\Xi^{-}}^{0}} 
          \ \int d^{3} k_{2}\  d^{3} k_{3}\ 
            \delta({\bf k}_{1} + {\bf k}_{2} + {\bf k}_{3}) \nonumber \\
    & &\qquad\times \delta( E - \frac{k_{1}^{2}}{2 m_{N}} 
                       - \frac{k_{2}^{2}}{2
                         m_{\Lambda}} - \frac{k_{3}^{2}}{2 m_{\Lambda}}) 
        \ |M ({\bf k}_{1},{\bf k}_{2},{\bf k}_{3};{\bf k}_{\Xi}^{0})|^{2} 
                              \ ,                    \label{eq:39}
\end{eqnarray}
where the $\delta$-functions maintain energy momentum conservation. 
The amplitude $|M|^{2}$ is given as the square of the sum of the 
diagrams in Fig.~\ref{fig.1}; {\it i.e.}, one obtains
\begin{equation}
|M|^{2} = \frac{1}{3}\,\sum_{S_{1}}\ 
           |A_{\ell_{N};\ell_{\Xi}} 
            + B_{\ell_{N};\ell_{\Xi}}
              + C_{{\ell}_{\Lambda};\ell_{\Xi}}|^{2}\ ,\label{eq:40}
\end{equation}
with $A_{\ell_{N};\ell_{\Xi}}$, $B_{\ell_{N};\ell_{\Xi}}$, and 
$C_{{\ell}_{\Lambda};\ell_{\Xi}}$ given in Eqs.~(\ref{eq:35}), 
(\ref{eq:36}), and (\ref{eq:37}). 
Because we are summing over the spin projections and integrating over the
momenta of the two identical $\Lambda$ hyperons, we have included a factor 
of $1/2!$ in Eq. (\ref{eq:39}). This factor is physically crucial as it 
avoids double counting of states, that afterall represent only interchanges 
of identical particles. The integral in Eq.~(\ref{eq:40}) can be 
reduced to a two-dimensional one which needs to be evaluated with some 
degree of accuracy~\cite{AA66,C96}.

\section{Numerical Results}\label{sec.3}

Having established that the neutron differential energy spectrum (NDES) is 
basically the square of the coherent sum of the diagrams in 
Fig.~\ref{fig.1}, we are now in a position to determine whether the reaction 
$\Xi^{-} d\rightarrow n\Lambda\Lambda$ is, in fact, a means to investigate 
the final state $\Lambda\Lambda$ interaction. To achieve this we must:  
(i)~Solve the coupled integral equations for the amplitudes $X_{N;\Xi}$, 
$Y_{N;\Xi}$, and $Y_{\Lambda;\Xi}$, {\it i.e.} Eqs.~(\ref{eq:22}), 
(\ref{eq:27}), (\ref{eq:29}), and (\ref{eq:32}), and then construct
the breakup amplitudes $A_{\ell_N;\ell_\Xi}$, $B_{\ell_N;\ell_\Xi}$,
and $C_{\ell_N;\ell_\Xi}$ defined in Eqs.~(\ref{eq:35}),
(\ref{eq:36}), and (\ref{eq:37}). 
(ii)~Establish the existence of an enhancement in the FSI 
region that is sensitive to the choice of the 
$\Lambda\Lambda$ interaction. 

\subsection{The input amplitudes}\label{sec.3.1}

The proposed reaction is initiated by $\Xi^{-}$ capture from an atomic 
$s$ orbit; as a result we consider a low energy reaction in which the 
two-body interactions are dominated by the $S$-wave contribution. 
Inclusion of the Coulomb interaction in the initial state to get the 
$\Xi^-$ in an atomic orbital turns a relatively simple three-body 
problem into a much more difficult numerical calculation due to the 
complex analytic structure of the Coulomb amplitude. Because we are 
predominantly interested in the final $\Lambda$--$\Lambda$ interaction, 
we have assumed that the initial $\Xi^{-}$ is incident on the deuteron 
at an energy of one MeV, which is below the deuteron breakup threshold. 
This allows us to neglect the Coulomb potential, since we don't have an 
initial bound state, and will enable us to check the numerical accuracy 
of our amplitudes with the help of three-body unitarity. 

Since we are assuming an $S$-wave interaction and considering an initial
state below the breakup threshold for $\Xi NN$, we may assume that all 
orbital angular momenta in the problem can be set to zero. As a result, 
the partial wave expansion for the elastic and rearrangement amplitudes 
reduces to
\begin{equation}
X_{\ell_{\alpha};\ell_{\beta}}({\bf k}_{\alpha};{\bf k}_{\beta}) = 
\frac{1}{4\pi}\ X_{\ell_{\alpha};\ell_{\beta}}(k_{\alpha};k_{\beta})
      \ .                                            \label{eq:41}
\end{equation}
With this limitation on the angular momentum, the coupled integral 
equations can be reduced to a set of one dimensional equations with 
the channel $\ell_\alpha$ labeled by the total spin and isospin 
$(S_{\alpha},t_{\alpha})$ of the pair $(\beta,\gamma)$. In 
Table~\ref{table.1} we present the nine channels 
included in the solution of our coupled integral equations for total 
spins and isospin $\frac{1}{2}$. Because the kernel of these equations 
have the standard moving singularities encountered in any three-body 
breakup problem, we have followed the same procedure implemented 
in the $n$--$d$ breakup reaction using the rotation-of-contour 
method~\cite{C71,CS71,C96}.
To test the accuracy of our numerical procedures in solving the
coupled integral equations and the construction of the breakup
amplitudes corresponding to the three diagrams in Fig.~\ref{fig.1}, 
we have made use of three-body unitarity to compare the imaginary
part of the elastic $\Xi$--$d$ amplitude to the total cross section. 
To calculate the total cross section for elastic $\Xi d$ scattering,
and in particular the imaginary part of the elastic amplitude, we also need
to solve the integral equation for the spin quartet, or 
${\cal S}=\frac{3}{2}$ case. In Table~\ref{table.2} we have the 
three-body channels for the ${\cal S}=\frac{3}{2}$ amplitudes.

To evaluate the NDES, we must  define the input two-body separable 
potential. From Tables~\ref{table.1} and \ref{table.2} it is clear that
we need the two-body interactions in the following channels:
\begin{eqnarray}
NN                 & : & ^1S_0,\ ^3S_1 \nonumber \\
\Xi N              & : & ^1S_0\ t=0,\quad ^3S_1\ t=0,1 \label{eq:42}\\
\Xi N-\Lambda\Lambda &: & ^1S_0      \ .  \nonumber
\end{eqnarray}
For the $NN$ potentials we have used the $S$-wave separable potentials
of Afnan and Gibson~\cite{AG90}, while the $\Xi N$ and 
$\Xi N$--$\Lambda\Lambda$ potentials are those constructed to give the 
same effective range parameters as the $SU(3)$ rotated Nijmegen model 
D\cite{NRS77} with soft core~\cite{C96,CAG97}.  Since we will examine 
the sensitivity of the FSI peak to the choice of the 
$\Lambda\Lambda$--$\Xi N$ interaction, we present in Table~\ref{table.3} 
the effective range parameters and the $\Lambda$--$\Lambda$ 
binding energy for the four potentials under consideration. Here we note 
that potentials SA and SB do not support a $\Lambda$--$\Lambda$ bound 
state, while potential SC1 and SC2 do support a bound state. In 
particular, the potential SB has a $\Lambda$--$\Lambda$ scattering length 
that is comparable to the $^1$S$_0$ $n$--$n$ scattering length; {\it i.e.}, 
it has a virtual state with a pole in the amplitude on the second energy 
sheet near the $\Lambda$--$\Lambda$ elastic threshold.

\subsection{Three-body unitarity}\label{sec.3.2}

Three-body unitarity for this reaction gives a relation between the 
imaginary part of the elastic $\Xi d$ amplitude and the total cross 
section at the corresponding energy; {\it i.e}, one has
\begin{equation}
Im\left[\,X_{\Xi;\Xi}\,\right] = -\frac{1}{16\pi^{3}}\,v_{\rm rel}\ 
                            \sigma_{\rm Tot}\ ,      \label{eq:43}
\end{equation}
where the relative velocity $v_{\rm rel}=k^{0}_{\Xi}/\mu_{\Xi d}$, 
and the total cross section is the sum of the elastic, breakup, and 
reaction cross sections, or
\begin{equation}
\sigma_{\rm Tot} = \sigma_{\rm el} + \sigma_{\rm B-up} 
                 + \sigma_{\rm reac}  \ .            \label{eq:44}
\end{equation}
We can now determine the total cross section $\sigma_{\rm Tot}$ from 
unitarity, while the elastic scattering cross section can be
determined by integrating the elastic differential cross section. 
This gives us, in the absence of a $\Lambda$--$\Lambda$ bound state, 
the total breakup cross section as
\begin{equation}
\sigma_{\rm B-up} = \sigma_{\rm Tot} - \sigma_{\rm el}
                           \ ,                     \label{eq:45}
\end{equation}
while in the presence of a $\Lambda$--$\Lambda$ bound state
\begin{equation}
\sigma_{\rm B-up} = \sigma_{\rm Tot} - \sigma_{\rm el}
                  - \sigma_{\rm reac}\ ,        \label{eq:46}
\end{equation}
with $\sigma_{\rm reac}$ being calculated by integrating the 
rearrangement differential cross section for $\Xi + d\rightarrow 
n + (\Lambda\Lambda)$. 

As a test of the numerical accuracy in the solution of the integral 
equation, we compare the results for the total breakup cross section as 
calculated from unitarity and the elastic and reaction cross section
({\it i.e.} Eq.~(\ref{eq:45}) or Eq.~(\ref{eq:46}) ), with the result 
of integrating the NDES over the neutron energy. In Table~\ref{table.4} we 
compare the results for these two methods of calculating the total 
breakup cross section. We have included results for two of 
the four potentials we will consider. Potential SB has no 
$\Lambda$--$\Lambda$ bound state, while potential SC1 has a bound state, 
and therefore a rearrangement cross section. The comparison is done at
an energy of 25.7~MeV above the $N\Lambda\Lambda$ threshold which is just below
the $\Xi NN$ threshold. The good agreement between the two methods used 
to construct the total breakup cross section is a clear indication that
the numerical procedure for calculating NDES is satisfactory. 
This is particularly ture when the unitarity approach for 
calculating  $\sigma_{\rm B-up}$ involves taking the difference
between the total and total elastic cross sections with both being 
an order of magnitude larger than the total breakup cross section.

\subsection{The neutron differential energy spectrum}\label{sec.3.3}

We are now in a position to examine the NDES for the different interactions.
In Fig.~\ref{fig.2} we present the neutron differential energy spectrum for 
the four potentials under consideration. Here we observe that for 
low energy neutrons all four potentials give basically the same shape for 
the cross section with the broad peak for low energy neutrons being
independent of the choice of potential for the $\Lambda\Lambda$--$\Xi N$ 
$^1$S$_0$ channel. On the other hand in the FSI region ({\it i.e.} for 
large neutron energies) the four potentials give different results
with potential SB having a small FSI peak. Considering
the similarity between the potential SB and the $^1$S$_0$ $n$--$n$ potential,
it is surprising that we don't see a large FSI peak 
similar to that observed in the $n$--$d$ breakup reaction. 
In the $n$--$d$ breakup, 
the enhancement in the cross section at the end of the proton spectrum
is a result of the fact that the energy available for the final 
$n$--$n$ interaction is small, and therefore the cross section is 
dominated by the pole in the two-body $n$--$n$ amplitude near zero 
energy which is known as the $^1$S$_0$ virtual state. The questions then are: 
(i)~ Why don't we see a similar FSI peak in 
the present reaction? 
(ii)~Why is the broad peak at the low energy part of the spectrum 
the same for the four potentials?

\begin{figure}
\vskip 0.5 cm
\centerline{\epsfig{figure=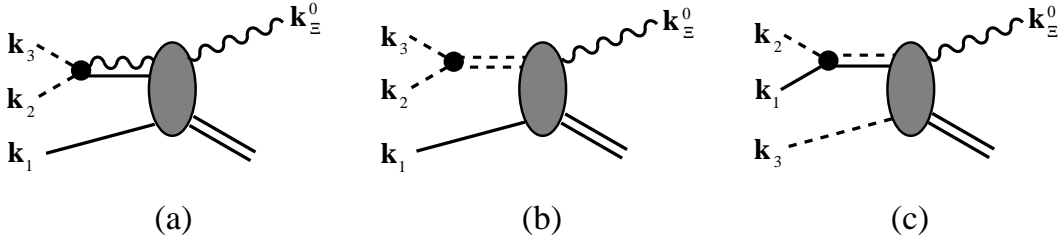,width=14cm}}
\vskip 0.5 cm
\caption{The three distinct contributions to the breakup amplitude, 
with (a) and (b) corresponding to the first term in Eq.~(\ref{eq:17}) 
and (c) corresponding to the last two terms in 
Eq.~(\ref{eq:17}).}\label{fig.1}
\end{figure}

\begin{figure}
\vskip 0.5 cm
\centerline{\epsfig{figure=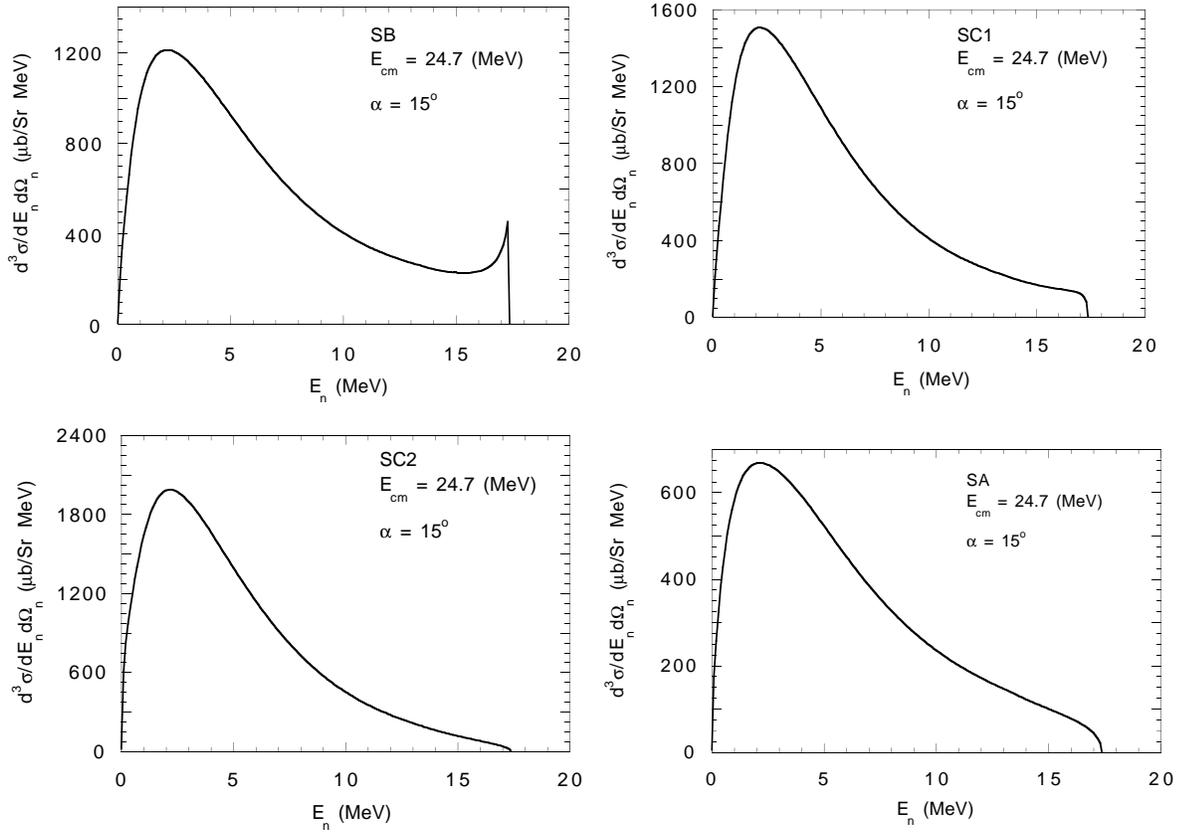,width=17cm}}
\vskip 0.5 cm
\caption{The NDES following the capture of $\Xi^{-}$ on the deuteron for 
the potentials SA, SB, SC1, and SC2.}\label{fig.2}
\end{figure}

\begin{figure}
\vskip 0.5 cm
\centerline{\epsfig{figure=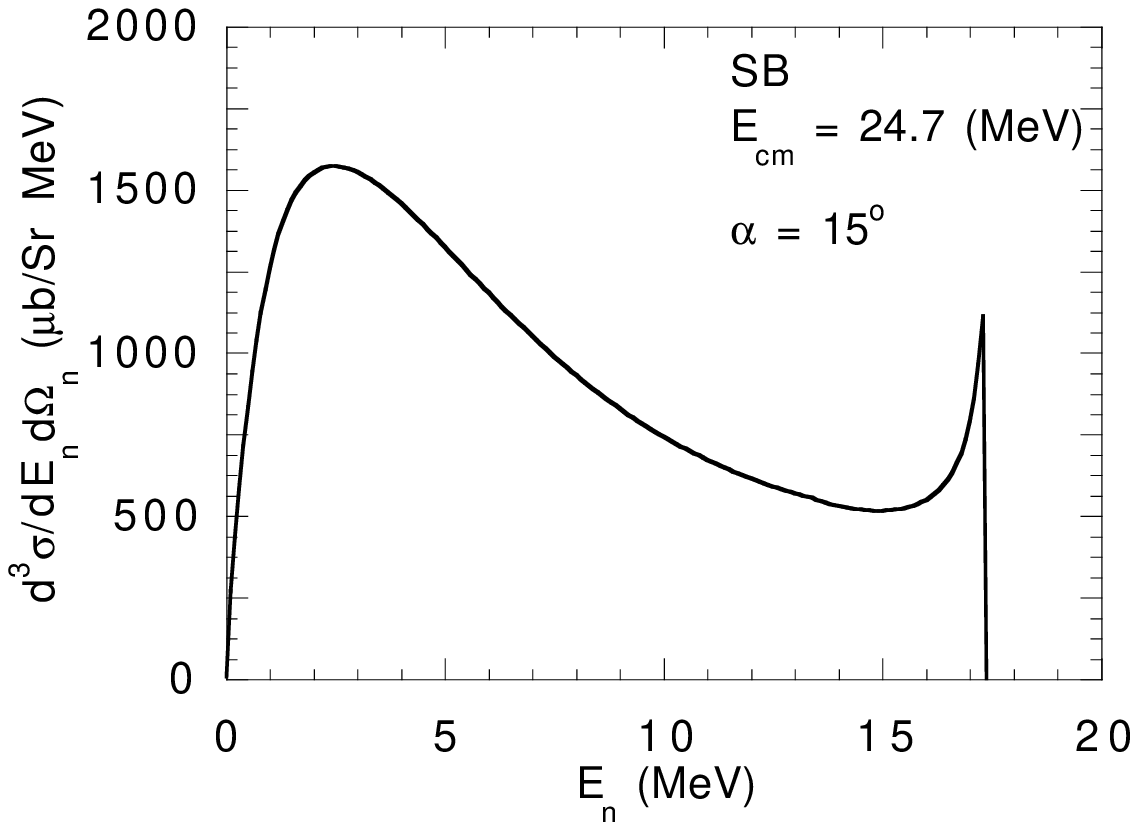,width=12cm}}
\vskip 0.5 cm
\caption{The NDES following the capture of $\Xi^{-}$ on the deuteron for 
the potentials SB including diagrams (a) and (c) of 
Fig.~\ref{fig.1}.}\label{fig.3}
\end{figure}

\begin{figure}
\vskip 0.5 cm
\centerline{\epsfig{figure=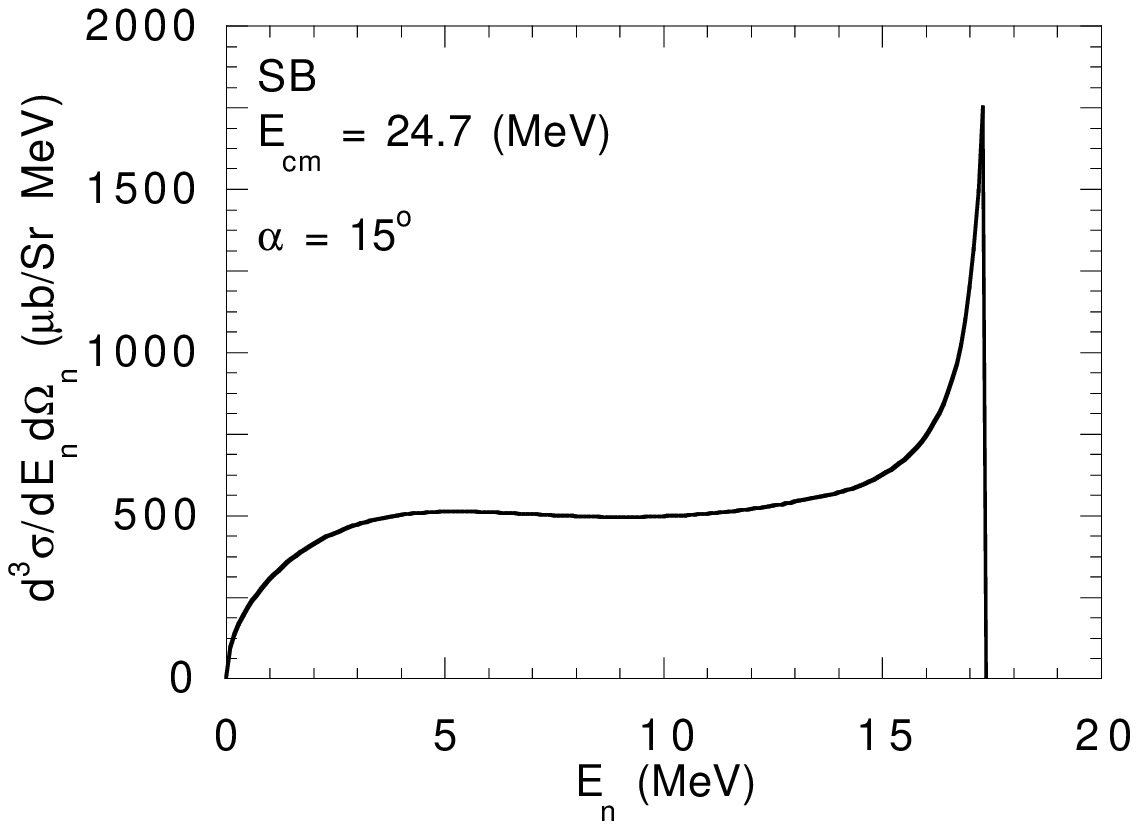,width=12cm}}
\vskip 0.5 cm
\caption{The NDES following the capture of $\Xi^{-}$ on the deuteron for 
the potential SB including diagrams (b) and (c) of 
Fig.~\ref{fig.1}.}\label{fig.4}
\end{figure}

To answer the first question, we consider the cross section for 
potential SB in more detail. Since the FSI peak 
should be the result of a pole near zero energy in the 
$\Lambda\Lambda$--$\Xi N$ $^{1}$S$_{0}$ amplitude, we expect a  
contribution to this FSI to come from diagrams (a)
and (b) of Fig.~\ref{fig.1}. We therefore consider the NDES for 
the following combination of diagrams:
(i)~The diagrams (a) and (c) of Fig.~\ref{fig.1}, see Fig.~\ref{fig.3}. 
(ii)~The diagrams (b) and (c) of Fig.~\ref{fig.1}, see Fig.~\ref{fig.4}. 
Here we observe that in both cases there is an enhancement in the
FSI as we originally expected. However the
magnitude of the FSI peak is considerably larger 
in Fig.~\ref{fig.4} than in Fig.~\ref{fig.3}. This suggests that 
we have constructive interference between diagrams (b) and (c), and
destructive interference between diagrams (a) and (c) of
Fig.~\ref{fig.1}. In this way we establish the fact that diagrams
(a) and (b) of Fig.~\ref{fig.1} are of opposite sign leading to a
cancellation between the two breakup amplitudes that provide the
enhancement due to the final state $\Lambda$--$\Lambda$ interaction.

\begin{figure}
\vskip 0.5 cm
\centerline{\epsfig{figure=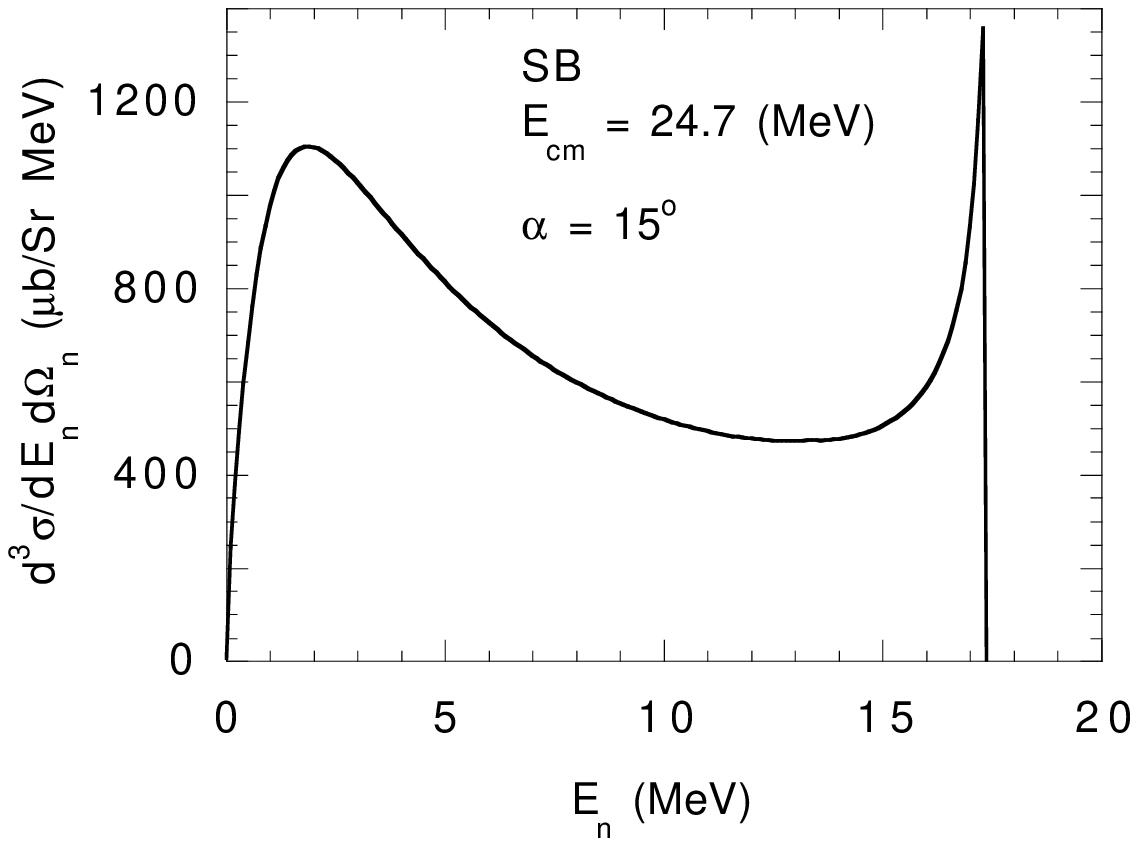,width=12cm}}
\vskip 0.5 cm
\caption{The NDES following the capture of $\Xi^{-}$ on the deuteron for 
the potentials SB including diagram (a)  of 
Fig.~\ref{fig.1}.}\label{fig.5}
\end{figure}

\begin{figure}
\vskip 0.5 cm
\centerline{\epsfig{figure=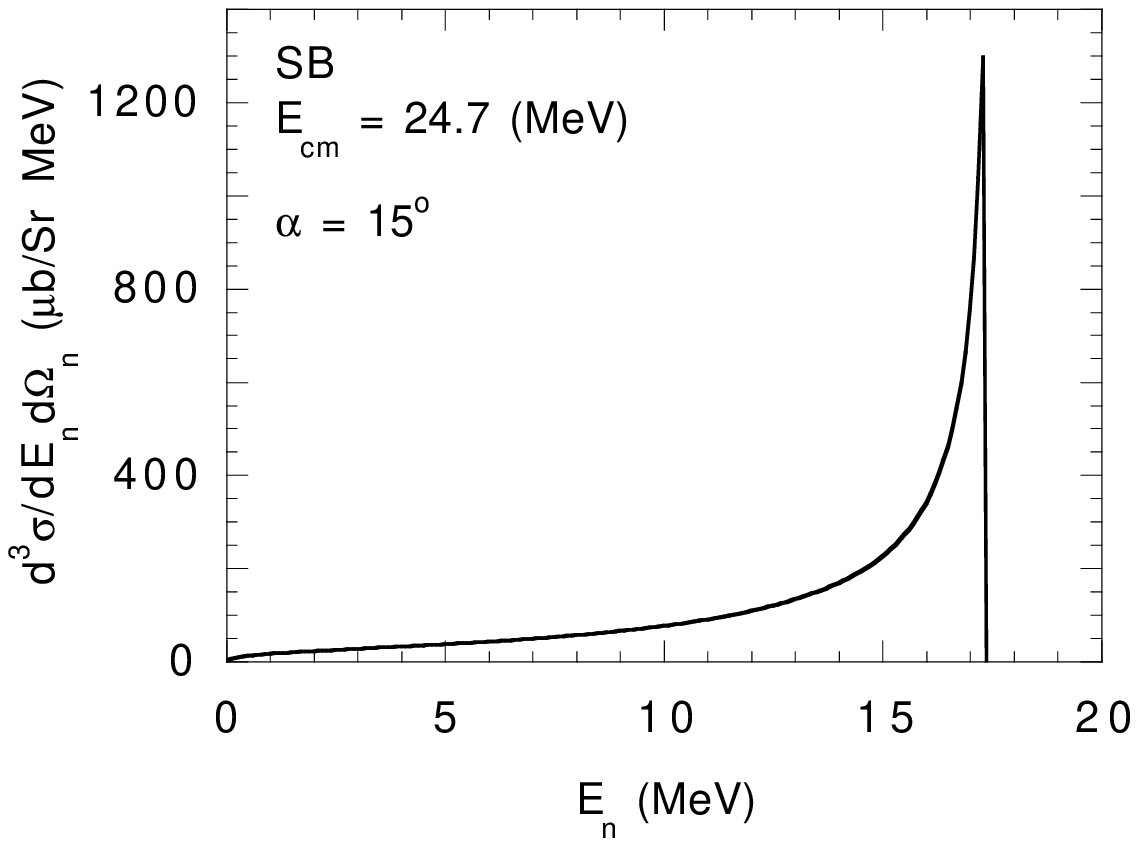,width=12cm}}
\vskip 0.5 cm
\caption{The NDES following the capture of $\Xi^{-}$ on the deuteron for 
the potentials SB including diagram (b)  of 
Fig.~\ref{fig.1}.}\label{fig.6}
\end{figure}

To establish the relative contribution of diagrams (a) and (b) of
Fig.~\ref{fig.1}, we present in Figs.~\ref{fig.5} and \ref{fig.6} the
NDES for diagrams (a) and (b) respectively. A comparison of the
magnitude of the FSI peak in these two diagrams
establishes the fact that the contribution of these two diagrams to the
total amplitude is the same. This is expected considering the fact
that it is the same pole that gives the FSI peak
in the two figures. From the above analysis we may conclude
that diagrams (a) and (b) almost cancel each other and the only
reason we have any FSI peak in the total amplitude
is a result of the fact that diagram (c) enhances diagram (b) relative
to diagram (a). The obvious difference between this reaction and 
that for the $n$--$d$ breakup is the fact that in this reaction we 
have a coupled channel problem for the FSI ({\it i.e.},
$\Lambda\Lambda$--$\Xi N$), while in the $n$--$d$ case we have only
one channel for the final $n$--$n$ interaction. Finally, we note that 
this cancellation does not depend on the sign of the coupling
$C_{\Lambda;\Xi}$ between the $\Lambda\Lambda$ and $\Xi N$ in the 
potential, but requires that the same interaction be used for the 
FSI as is used in generating the inelastic and breakup amplitudes. 
To understand this we note that a change in the sign of
$C_{\Lambda;\Xi}$ changes the sign of $\tau_{\Lambda\Lambda;\Xi N}$
but not the sign of $\tau_{\Lambda\Lambda;\Lambda\Lambda}$, and since
each term in the multiple scattering series for each of the three 
amplitude has an odd number of $\tau_{\Lambda\Lambda;\Xi N}$, the sign
of all three amplitudes changes under the substitution 
\[
C_{\Lambda;\Xi} \rightarrow -\ C_{\Lambda;\Xi}\ .
\]
As a result the cross section does not give us any insight into the
sign of the coupling between the $\Lambda\Lambda$ and $\Xi N$
channels. 

\begin{figure}
\vskip 0.5 cm
\centerline{\epsfig{figure=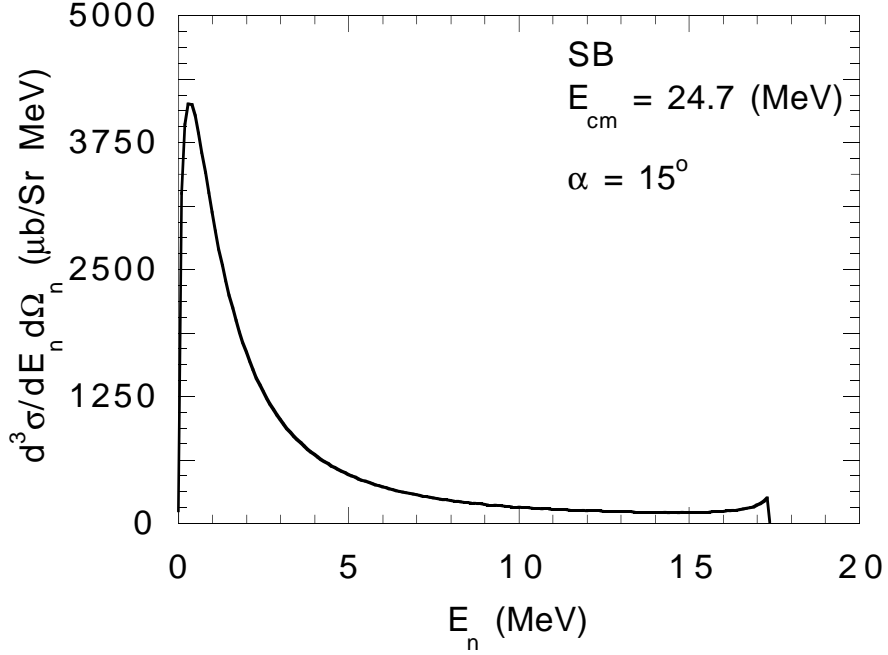,width=12cm}}
\vskip 0.5 cm
\caption{The NDES following the capture of $\Xi^{-}$ on the deuteron for 
the potentials SB including diagram (a) of 
Fig.~\ref{fig.1} to lowest order, {\it i.e} 
$g_{N}\,\tau_{\Lambda\Lambda;\Xi N}\,Z_{N;\Xi}$.}\label{fig.7}
\end{figure}

We now turn to the broad peak in the NDES at the low end of the 
neutron spectrum. Here, we first observe from Fig.~\ref{fig.6} that 
diagram (b) does not contribute to this peak. The fact that it is not 
present in the NDES resulting from diagrams (b) and (c), see
Fig.~\ref{fig.4}, suggests that the main contribution to the broad peak
at the low energy end of the neutron spectrum is due to diagram (a), 
which in lowest order is given by
\[
g_{N}(p_1)\,\tau_{\Lambda\Lambda;\Xi N}\,X_{N,\Xi}\approx
g_{N}(p_1)\,\tau_{\Lambda\Lambda;\Xi N}\,Z_{N,\Xi}\  ,
\]
and corresponds to the $\Xi^-$ interacting with the proton in the deuteron 
to generate two $\Lambda$ hyperons. In this case the neutron spectrum is 
determined by the momentum distribution in the deuteron. To illustrate 
this we present in Fig.~\ref{fig.7} the lowest order contribution from 
diagram (a) of Fig.~\ref{fig.1} to the NDES in which the FSI 
is almost non-existent. However, as we add the higher order contributions 
in the (divergent) multiple scattering for $X_{N;\Xi}$ to the NDES, the 
low energy peak is reduced in magnitude and is converted from a sharp peak 
to the broad peak we found in the full calculation. From this we may 
conclude that the broad peak in the neutron spectrum is to a large extent 
determined by the momentum distribution in the deuteron, and since the same 
deuteron wave function was used to generate the results for the four 
$\Lambda\Lambda$--$\Xi N$ potentials in Fig.~\ref{fig.2}, the shape of 
the peak is basically the same. The difference in magnitude is due to 
the difference in the $\Lambda\Lambda$-$\Xi N$ potentials used.

The fact that the FSI peak is suppressed as a result of the
cancellation between the diagrams (a) and (b) of Fig.~\ref{fig.1},
suggests that the height of the peak might be sensitive to the coupling
between the $\Lambda$--$\Lambda$ and $\Xi$--$N$ channels, since the 
absence of such coupling in the FSI would have given us the result in 
Fig.~\ref{fig.4} in which the NDES is dominated by the FSI peak. 
The potential SB is based on a separable approximation to a one 
boson exchange potential (OBE) in which the long range part was 
determined by the $SU(3)$ rotation of Nijmegen model D potential. The 
short range part of the interaction was adjusted in such a way that 
the $SU(3)$ rotated OBE potential was not altered outside of 
0.8~fm\cite{C96,CAG97}. Because the long range part of the coupling 
potential is determined by the $K$- and $K^*$-exchanges which in turn is 
determined by $SU(3)$ except for the masses of the mesons and baryons,
we expect the degree of cancellation between diagrams (a) and (b) may 
be determined by $SU(3)$. In an attempt to test this idea, we scaled 
the coupling between the $\Lambda$--$\Lambda$ and $\Xi$--$N$ 
channels; {\it i.e.}, we assumed
\[
C_{\Lambda\Xi} \rightarrow R \ C_{\Lambda\Xi} \ ,
\]
where $C_{\Lambda\Xi}$ is the strength of the coupling between the
$\Lambda$--$\Lambda$, and the $\Xi$--$N$ channels in the separable
potential. Here, we chose the scale factor $R=0.5, 0.75, 1.0$, and 
$1.25$. In this way we could investigate changes in the FSI peak with 
variation in the coupling between the $\Lambda$--$\Lambda$ channel 
and the $\Xi$--$N$ channel. To maintain the position of the virtual 
pole in the amplitudes fixed, we adjusted the other parameters of 
the potential to retain a scattering length 
$a_{\Lambda\Lambda}\approx 21$~fm.

\begin{figure}
\vskip 0.4 cm
\centerline{\epsfig{figure=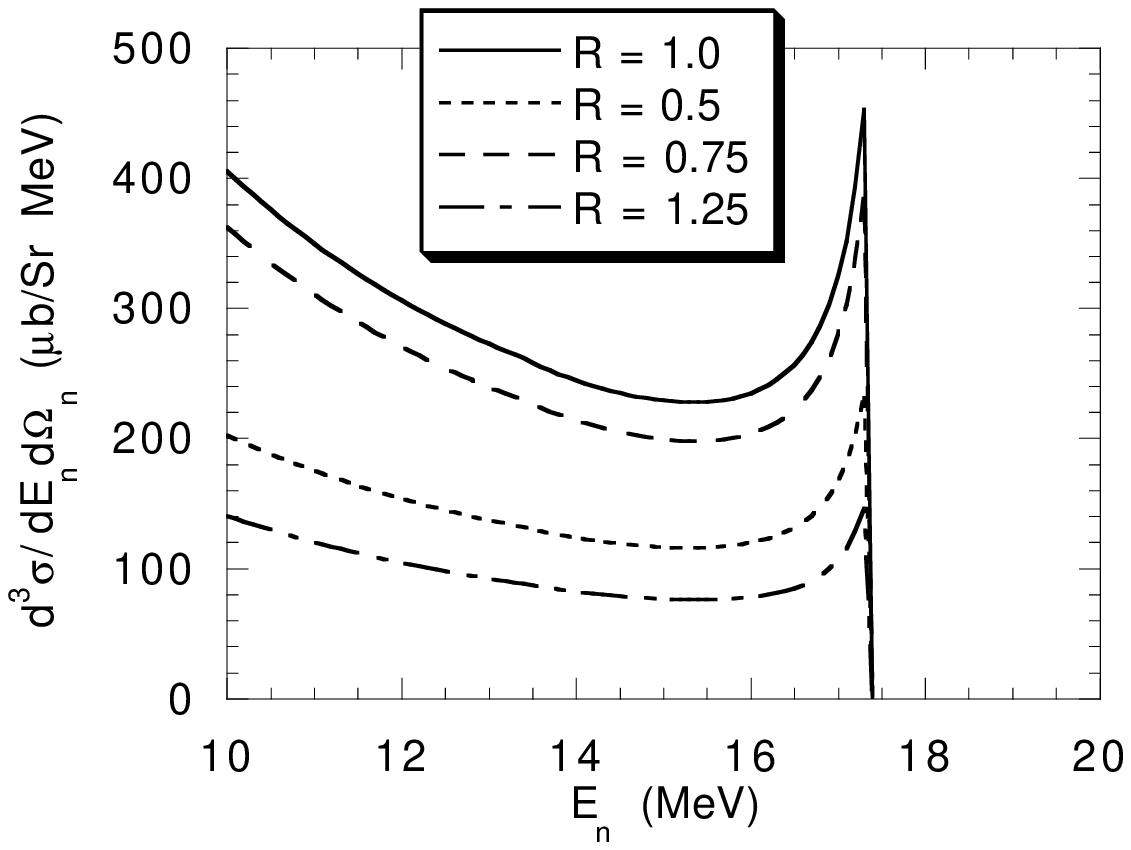,width=12cm}}
\vskip 0.4 cm
\caption{The NDES following the capture of $\Xi^{-}$ on the deuteron,  
over the FSI region, for a series of interactions in which the
strength of the coupling between the $\Lambda$--$\Lambda$ and
$\Xi$--$N$ is modified by multiplying by a factor $R$, {\it i.e.}
$C_{\Lambda\Xi}\rightarrow R\,C_{\Lambda\Xi}$ with $R=0.5, 0.75, 1.0$,
and $1.25$.}\label{fig.8}
\end{figure}

In Fig.~\ref{fig.8} we present the NDES in the FSI region for different 
values of $R$. We note here that the $R=1$ case gives the largest FSI 
peak, with the other values of $R$ invariably reducing the magnitude of 
the FSI peak. This suggests that a measurement of the NDES could
give more than the $\Lambda$--$\Lambda$ scattering length in that the
height of the peak could shed some light on the strength of the
coupling between the $\Lambda$--$\Lambda$ and $\Xi$--$N$ channels and,
in this way, could test the degree of violation of $SU(3)$ symmetry
in the two baryon system. If the suggestion of Dover and
Feshbach\cite{DF90} that there is much less dynamic $SU(3)$ symmetry 
breaking in the $\Lambda\Lambda\rightarrow\Xi N$ than in
$\Lambda\Lambda\rightarrow\Lambda\Lambda$, then the height of the FSI 
peak might be used to explore this question.


\section{Conclusions}\label{sec.4}

The main motivation for examining the reaction 
$\Xi^-d\rightarrow n\Lambda\Lambda$ was to 
determine whether this reaction might provide some
insight into the $\Lambda$--$\Lambda$ interaction at low energies and
possibly give a measure of the scattering length. In this way we can
establish whether the $\Lambda$--$\Lambda$ and the $n$--$n$ belong to 
the same flavor multiplet as predicted by $SU(3)$. At the same time if
this reaction is to be used to establish the existence of the $H$
dibaryon, then we should establish some quantitative measure of what 
the cross section would be in the absence of the $H$. In particular, 
we need to establish which features of the cross section 
can be reproduced without the introduction of the $H$ particle. 

From the results of the above calculation we have established that the 
neutron differential energy spectrum does, in fact, give a FSI 
peak in the event that the $\Lambda$--$\Lambda$
amplitude has a pole on the second Riemann sheet of the energy plane
near the $\Lambda$--$\Lambda$ threshold. Thus, experimental
observation of a FSI peak is a strong signature
that the $\Lambda$--$\Lambda$ interaction is, in fact, similar to the
$n$--$n$ interaction as suggested by flavor $SU(3)$. In contrast to
the $n$--$n$ interaction, the $\Lambda$--$\Lambda$ interaction is part
of the coupled channel $\Lambda\Lambda$--$\Xi N$, and we have
established that in the case when the FSI is a
coupled channel problem, then there are two distinct diagrams that
contribute to the FSI peak. In this case the
two diagrams (a) and (b) of Fig.~\ref{fig.1} almost cancel each other
resulting in a suppression of the FSI peak. In fact the background
contribution from the diagram (c) in Fig.~\ref{fig.1} effectively
preserves the final state interaction by interfering constructively
with diagram (b) and destructively with digram (a). This cancellation
is a unique feature of the coupled channel nature of the FSI.

The presence of the interference between diagrams (a) and (b) of
Fig.~\ref{fig.1} makes this reaction a possible tool to measure the
strength of the coupling between the $\Lambda$--$\Lambda$ and
$\Xi$--$N$ channels. Since this coupling is determined in the OBE
model by the $K$- and $K^*$-exchanges whose coupling to the baryon is
determined by $SU(3)$, we expect the magnitude of the FSI peak to be a
candidate for the determination of $SU(3)$ violation in the two baryon
system.

Finally, we observe that the low energy part of the NDES has a broad
peak which is largely determined by the momentum distribution in the
target deuteron. Although this is not the ideal reaction to examine
the deuteron wave function, the fact that we have a major cancellation
between the three diagrams that contribute to the FSI suggests that the
FSI peak might be sensitive to the choice of deuteron wave
function. In the present investigation we have used the simplest
possible $^3$S$_1$ deuteron, and a more realistic wave function for
the deuteron ({\it e.g.} one with short range repulsion) could give a
different result for the magnitude of the FSI peak. The magnitude of
the FSI peak could also be sensitive to the choice of the $\Xi$--$N$ and
the $\Lambda\Lambda$--$\Xi N$ interactions. For the present
investigation we have chosen to use simple separable potentials that
give the same effective range parameters as the OBE potentials. The
existence of experimental data would be required to justify the extension 
of the present results to include more realistic two-body input.


\acknowledgments

S.B.C.\ and I.R.A.\ would like to thank the Australian Research Council for 
their support.  The work of B.F.G.\ was performed under the auspices 
of the U.S.\ Department of Energy.  I.R.A. and B.F.G. thank the Institute for 
Nuclear Theory at the Univ.\ of Washington for its hospitality during
the initial discussions of this problem with Carl Dover and the Institut fuer 
Kernphysik of the Forschhungszentrum Juelich for its hospitality when 
this manuscript was drafted.

\appendix

\section{The kernel of the AGS equations}\label{app.1}
To solve the coupled integral equations for the elastic and
rearrangement amplitudes $X_{\ell_\alpha;\ell_\beta}$ and
$Y_{\ell_\alpha;\ell_\beta}$, we need to define the kernel or the Born
amplitudes $Z_{\ell_\alpha;\ell_\beta}$. Since the reaction under
consideration involves two Hilbert spaces, {\it i.e} the $\Xi NN$ and
$\Lambda\Lambda N$, and the Born term does not couple these two
spaces, we can define the Born term for each space separately. Thus, in
the $\Xi NN$ Hilbert space we have the nucleon exchange amplitude
$Z_{\ell_N;\ell_\Xi}$ and the $\Xi$ exchange amplitude
$Z_{\ell_N;\ell_N}$. To evaluate these amplitudes we need to relate
them to the corresponding amplitudes in which the the initial and final
state are related by the cyclic permutation of the particle label for
which we have a standard expression~\cite{AG90}. Thus the $N$-exchange is
given by:
\begin{eqnarray}
Z_{\ell_N;\ell_\Xi} &=& \left[Z_{\ell_\Xi;\ell_N}\right]^\dag 
 \equiv \frac{1}{\sqrt{2}}\,\left[\,Z_{N_1;\Xi_3} 
                    - Z_{N_2;\Xi_3}\,\right] \nonumber \\
                    &=& \sqrt{2}\,Z_{1;3} 
         \equiv\sqrt{2}\,\langle(23)1|G_0|(12)3\rangle \ ,\label{eq:a1}
\end{eqnarray}
where particles 1 and 2 are the nucleon, while particle 3 is the
$\Xi$. The $\Xi$ exchange Born amplitude can be written in a similar
manner as:
\begin{eqnarray}
Z_{\ell_N;\ell_N} &\equiv& -\,Z_{N_1;N_2} = (-1)^{P+1}\ Z_{1;2}
                               \nonumber \\
                  &\equiv& (-1)^{P+1}\ \langle(23)1|G_0|(31)2\rangle
                                \ .                     \label{eq:a2}
\end{eqnarray}
Here the phase $P$ results from the exchange of the coordinates of
particles 1 and 3 in the coupling coefficient and is given by $P=s_1
+s_3 - S_2 +\tau_1 + \tau_3-t_2$, where $s_i$ and $\tau_i$ are the
spin and isospin of particle $i$, which in this case is $\frac{1}{2}$,
while $S_\alpha$ and $t_\alpha$ are the total spin and isospin of the
pair $(\beta\gamma)$.

On the other hand, in the $\Lambda\Lambda N$ Hilbert space, we need the
$N$-exchange Born amplitude $Z_{\ell_N;\ell_\Lambda}$ and the
$\Lambda$ exchange amplitude $Z_{\ell_\Lambda;\ell_\Lambda}$. These are
given in terms of the cyclicly defined Born amplitudes as:
\begin{eqnarray}
Z_{\ell_N;\ell_\Lambda} &=& \sqrt{2}\,Z_{N_1;\Lambda_2} 
                         = \sqrt{2}\,Z_{1;2} \nonumber \\
        &\equiv& \sqrt{2}\ \langle(23)1|G_0|(31)2\rangle\ ,\label{eq:a3}
\end{eqnarray}
and
\begin{eqnarray}
Z_{\ell_\Lambda;\ell_\Lambda} &\equiv& -\,Z_{\Lambda_3;\Lambda_2}
               = (-1)^{P'+1}\ Z_{2;3}\nonumber \\
            &\equiv& (-1)^{P'+1} \ \langle(31)2|G_0|(12)3\rangle
                                      \ .        \label{eq:a4}
\end{eqnarray}
In this case the the phase $P'$ results from the exchange of the
labels on particles 1 and 2, and is given by $P'= s_1 + s_2 - S_3 +
\tau_1 + \tau_2 -t_3$. The cyclic order Born amplitudes can 
in general be partial wave expanded for any three particles with specific
spin and isospin~\cite{AG90,AT77}.

\newpage


\newpage

\begin{table}
\caption{ Three-body channels allowed for the $\cal S$ = $\frac{1}{2}$, $T$ = 
$\frac{1}{2}$ configuration. \label{table.1}}   
\vspace{12pt}
\begin{tabular}{c|ccc} 
            & Channel         & $t_{\alpha}$ & $S_{\alpha}$ \\ \hline
$\Xi(NN)$:  & & & \\
            & NN ($ ^{3}$S$_{1}$)          & 0 & 1 \\
            & NN ($ ^{1}$S$_{0}$)          & 1 & 0 \\ \hline
$N(\Xi N)$: & & & \\ 
            & $\Xi N$ ($ ^{1}$S$_{0}$)     & 0 & 0  \\
            & $\Xi N$ ($ ^{1}$S$_{0}$)     & 1 & 0  \\
            & $\Xi N$ ($ ^{3}$S$_{1}$)     & 0 & 1  \\
            & $\Xi N$ ($ ^{3}$S$_{1}$)     & 1 & 1  \\ \hline
$N (\Lambda \Lambda)$: & & & \\
            & $\Lambda\Lambda$  ($ ^{1}$S$_{0}$) & 0 & 0  \\ \hline
$ \Lambda (N \Lambda)$: & & & \\
            & $\Lambda N$ ($ ^{1}$S$_{0}$) & $\frac{1}{2}$ & 0 \\
            & $\Lambda N$ ($ ^{3}$S$_{1}$) & $\frac{1}{2}$ & 1  
\end{tabular}
\vskip 1 cm
\caption{ Three-body channels allowed for the $\cal S$ = $\frac{3}{2}$, $T$ = 
$\frac{1}{2}$ configuration. \label{table.2} }
\vspace{12pt}
\begin{tabular}{c|rrr} 
 & Channel & $t_{\alpha}$ & $S_{\alpha}$ \\ \cline{1-4}
$\Xi (NN)$: & & & \\
            & NN ($ ^{3}$S$_{1}$) & 0 & 1 \\ \cline{1-4}
$N (\Xi N)$: & & & \\ 
             & $\Xi N$ ($ ^{3}$S$_{1}$) & 0 & 1  \\
             & $\Xi N$ ($ ^{3}$S$_{1}$) & 1 & 1  \\ 
\end{tabular}
\end{table}


\begin{table}
\centering
\caption{{The effective range parameters for the $\Lambda\Lambda$--$\Xi N$ 
 coupled channel separable potentials in the $^{1}S_{0}$ partial wave.}
\label{table.3} }    
\vspace{12pt}
\begin{tabular}{lccccc} 
  Pot. & $a_{\Lambda\Lambda}$~(fm) & $r_{\Lambda\Lambda}$~(fm) & 
       $a_{\Xi N}$~(fm) &  $r_{\Xi N}$~(fm) & B.E. (MeV) \\ \hline
    
    SA   & -1.90 & 3.33 & -2.08-0.81i & 3.44-0.22i & UB \\
      
    SB   & -21.0 & 2.54 & -2.07-6.52i  & 2.62-0.15i & UB \\

    SC1   & 7.84 & 1.48 & 3.05-5.28i & 1.45+0.074i & 0.71 \\
        
    SC2   & 3.36 & 1.0 & 3.35-2.50i  & 1.83-0.10i & 4.73 \\ 
\end{tabular}
\end{table}

\begin{table}
\caption{ The total breakup cross section as calculated via unitarity
($\sigma_{\rm B-up}^{\rm u}$) and by integrating the NDES 
($\sigma_{\rm B-up}^{\rm d}$) at an energy of 25.7~MeV. \label{table.4}}
\begin{tabular}{c|cc}
$\Lambda\Lambda-\Xi N$ & $\sigma_{\rm B-up}^{\rm u}$ 
                                       & $\sigma_{\rm B-up}^{\rm d}$ \\ \hline
SB    & 83.32   & 83.77  \\
SC1   &111.69   &111.69  \\
\end{tabular}
\end{table}


\begin{references}
\bibitem{foot1} We have used isospin in our formulation, and the projection
                of the correct initial, and therefore, final state is
                achieved by taking isospin projection of $-1/2$.
\end{references}
\end{document}